\documentclass[fleqn,12pt,twoside]{article}
\usepackage[headings]{espcrc1}

\readRCS
$Id: espcrc1.tex,v 1.2 2004/02/24 11:22:11 spepping Exp $
\usepackage{graphicx}
\usepackage[figuresright]{rotating}

\newcommand{\AmS}{{\protect\the\textfont2
  A\kern-.1667em\lower.5ex\hbox{M}\kern-.125emS}}

\title{Reaction mechanisms for weakly-bound, stable nuclei and unstable, 
halo nuclei on medium-mass targets}

\author{C. Beck\address[IPHC]{Institut Pluridisciplinaire Hubert Curien, 
        UMR 7178, CNRS-IN2P3 and Universit\'{e} de Strasbourg, 
        B.P. 28, F-67037 Strasbourg Cedex 2, France},
	N. Rowley\addressmark[IPHC]\thanks{Institut de Physique Nucl\'eaire,
	and Universit\'e de Paris Sud, F-91406 Orsay Cedex, France},
	P. Papka\addressmark[IPHC]\thanks{Department of Physics, University of
	Stellenbosch, Marieland, Stellenbosch 7602, South Africa}
	S. Courtin\addressmark[IPHC],
	M. Rousseau\addressmark[IPHC],
        F.A. Souza\address[USP]{Departamento de Fisica Nuclear, 
	Universidade de S\~ao Paulo, SP, Brazil},
	N. Carlin\addressmark[USP], 
	R. Liguori Neto\addressmark[USP],
	M.M. de Moura\addressmark[USP],
	M.G. Del Santo\addressmark[USP],
	A.A.P. Suaide\addressmark[USP]
	M.G. Munhoz\addressmark[USP],
	E.M. Szanto\addressmark[USP],
	A. Szanto de Toledo\addressmark[USP],
	N. Keeley\address[Warsaw]{The Andrzej Soltan Institute for Nuclear
	Studies, Warsaw, Poland},
	A. Diaz-Torres\address[Surrey]{Department of Physics, University of 
	Surrey, Guildford, UK}, 
	K. Hagino\address[TOHOKU]{Department of Physics, Tohoku University,
	Sendai, Japan}.
		}

\runtitle{Reaction mechanisms for weakly-bound, stable nuclei and unstable, 
halo nuclei
}
\runauthor{C. Beck}

\begin{document}

% typeset front matter
\maketitle

\begin{abstract}
An experimental overview of reactions induced by the stable, but weakly-bound 
nuclei $^{6}$Li, $^{7}$Li and $^{9}$Be, and by the exotic, halo nuclei 
$^{6}$He, $^{8}$B, $^{11}$Be and $^{17}$F on medium-mass targets, such 
as $^{58}$Ni, $^{59}$Co or $^{64}$Zn, is presented. Existing data on
elastic scattering, total reaction cross sections, fusion processes,
breakup and transfer channels are discussed in the framework of a CDCC 
approach taking into account the breakup degree of freedom.
\end{abstract}

\section{INTRODUCTION}

In reactions induced by stable, but weakly-bound nuclei and by 
exotic (unstable), halo nuclei, the influence on the fusion process of 
coupling to both collective degrees of freedom and transfer/breakup 
channels is a key point in understanding the dynamics of many-body quantum 
systems~\cite{Beck07a}. Due to their very weak binding energies, the ``halo" 
(a diffuse cloud of neutrons for $^{6}$He or an extended spatial distribution 
for the loosely-bound proton in~$^{8}$B) will lead to larger total reaction 
(and fusion) cross sections at sub-barrier energies when compared with 
predictions of one-dimensional barrier penetration models 
\cite{Beck07a,Aguilera09}. This enhancement is well understood in terms 
of the dynamical processes arising from strong couplings to collective 
inelastic excitations (such as soft-dipole resonances) of the target and 
projectile. However, in the case of reactions where at least one of the 
colliding nuclei has a sufficiently-low binding energy for breakup to become 
a competitive process, conflicting conclusions have been reported 
\cite{Beck07a,Aguilera09,Chatterjee08,Pietro04}. Recent studies with 
radioactive ion beams (RIB) indicate that the halo nature of $^{6}$He, for 
instance, does not enhance the fusion probability as much as anticipated. 
Instead, the prominent role of one- and two-neutron transfers in 
$^{6}$He-induced fusion
reactions~\cite{Chatterjee08,Bychowski04,DeYoung04,Penion06,Escrig07} 
has been definitively demonstrated. On the other hand, the effect of 
non-conventional transfer or stripping processes (see e.g.~\cite{Beck07a})
appears to be less significant for stable, weakly-bound projectiles.

\section{EXPERIMENTAL OVERVIEW}

\begin{figure}[htb]
\begin{minipage}[t]{80mm}
\includegraphics[width=79mm]{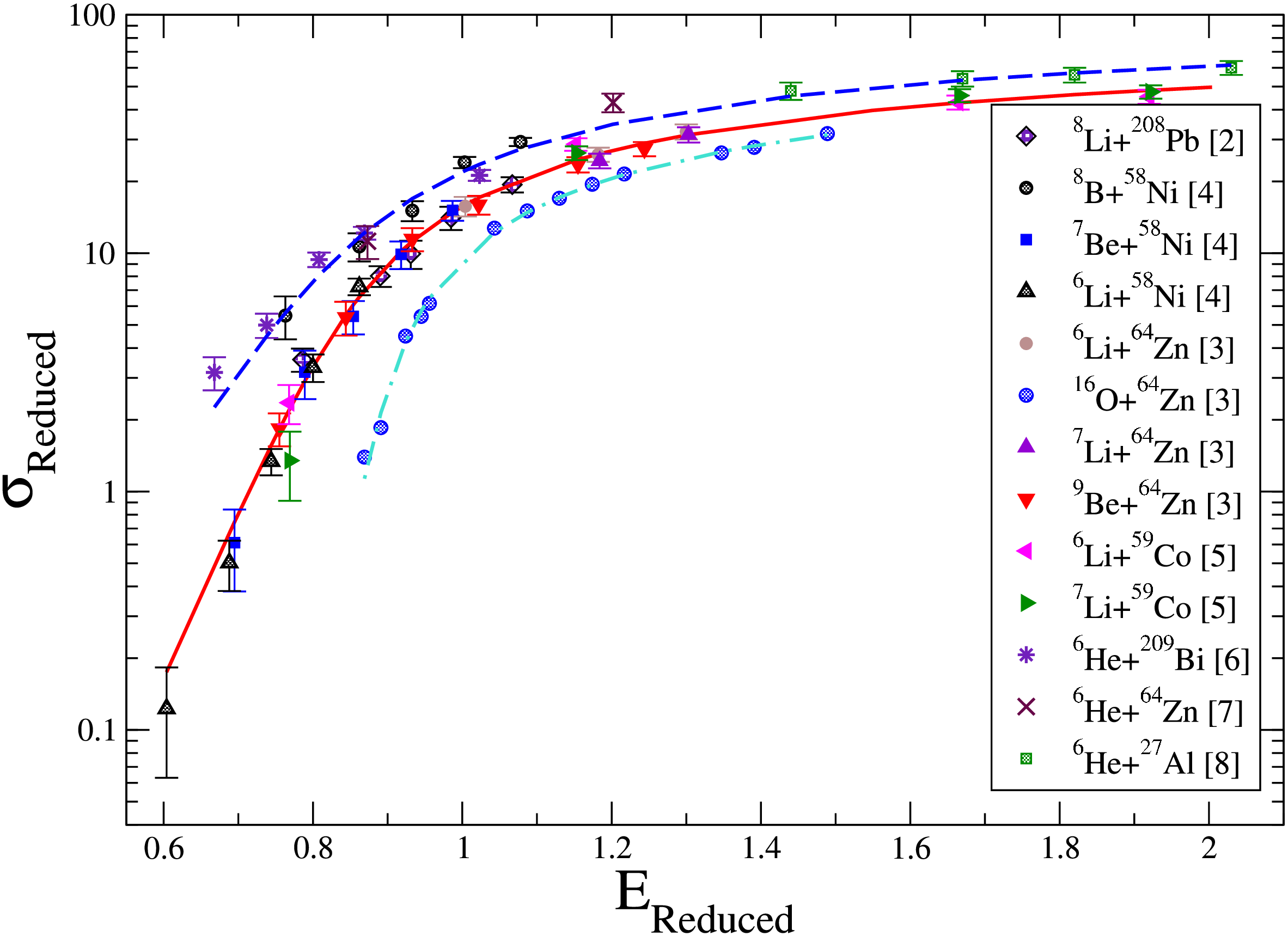}
\caption{(Color online) Reduced total reaction cross sections for a number of 
systems taken from the literature 
\cite{Aguilera09,Pietro04,Kolata02,Aguilera01,Benjamin07,Gomes05,Beck07b}. 
``Reduced" Energies and ``reduced" cross sections were extracted following 
procedures proposed in Refs. ~\cite{Aguilera09,Gomes05}. The figure was adapted 
from Ref.\cite{Kolata09}. }
\label{fig.1}
\end{minipage}
\hspace{\fill}
\begin{minipage}[t]{75mm}
\includegraphics[width=74mm]{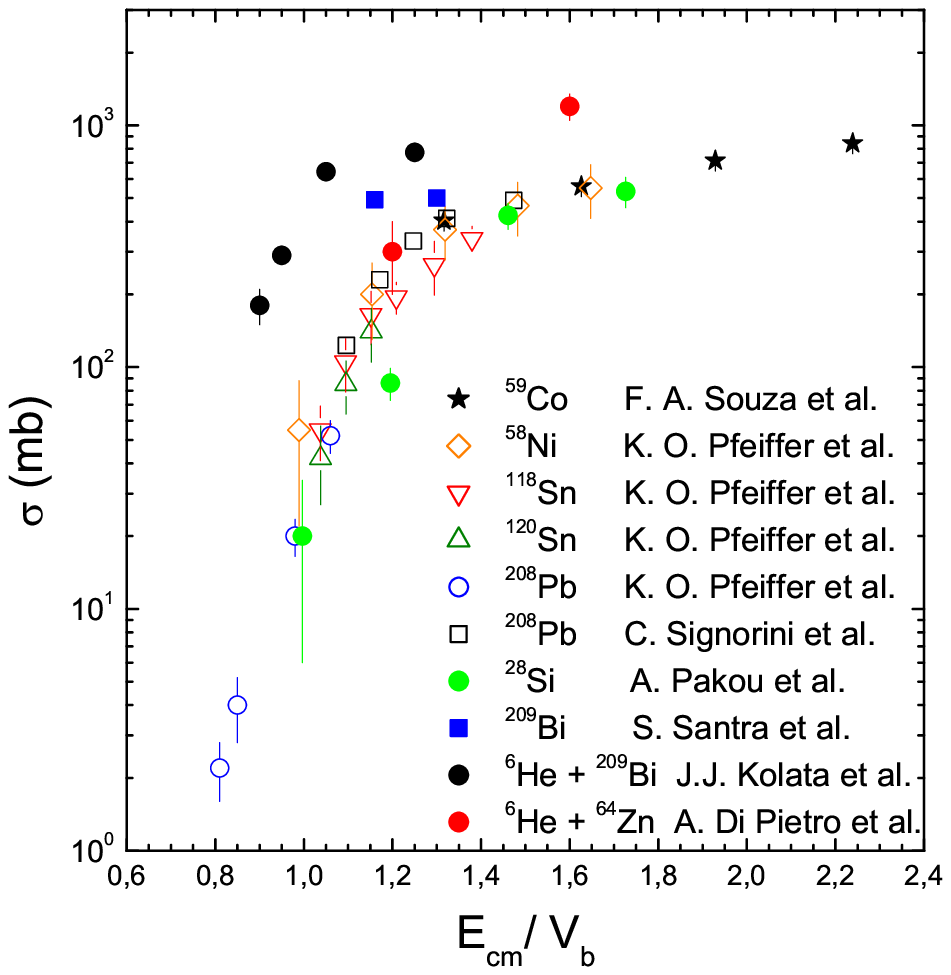}
\caption{(Color online) Total $\alpha$-production cross sections for a number 
of systems involving $^6$Li
\cite{Pfeiffer73,Signorini03,Beck03,Pakou03,Souza09,Santra09} and $^6$He 
projectiles~\cite{Pietro04,Kolata98}. In particular, the data corresponding
to the $^{6}$Li+$^{59}$Co reaction, and plotted as full stars, have been 
obtained from Ref.~\cite{Souza09}).}
\label{fig.2}

\end{minipage}
\end{figure}

Several experiments involving stable, weakly-bound projectiles such as 
$^{6}$Li, $^{7}$Li and $^{9}$Be on medium-mass targets have been
undertaken in the recent past~\cite{Beck07a,Gomes05,Beck03,Pakou03,Souza09}. 
The main results are summarized in Figs. 1 and 2 along with selected
experimental data obtained for $^{6}$He, $^{8}$Li, $^{8}$B, $^{7,11}$Be and 
$^{17}$F exotic beams 
\cite{Aguilera09,Chatterjee08,Pietro04,Kolata02,Aguilera01,Benjamin07,Mazzocco09,Pietro09}. 
Of particular interest are the very large total reaction cross sections 
observed in Fig.~1 for the ``halo" systems $^{6}$He+$^{64}$Zn~\cite{Pietro04}, 
$^{8}$B+$^{58}$Ni \cite{Aguilera09} and $^{6}$He+$^{209}$Bi~\cite{Kolata98} 
(upper red curve) in comparison with weakly-bound ``normal" systems (lower 
red curve). Please note that the data points for $^{6}$He+$^{208}$Pb 
\cite{Sanchez08} and $^{6}$He+$^{197}$Au \cite{Kakuee03} total reaction cross 
sections, being compatible with those measured for $^{6}$He+$^{209}$Bi
\cite{Kolata98}, were not plotted in Fig.~1.

The two classes of projectile yield two distinct curves for the 
``reduced" cross sections as a function of the ``reduced" energy (as defined 
in Refs.\cite{Aguilera09,Gomes05,Kolata09}). Both the ``halo" systems (a 
weakly bound proton is still confined by the Coulomb barrier and thus gives 
less than a halo) and the stable weakly-bound systems have cross sections 
which lie well above the systematic behavior defined by the very tightly-bound 
systems $^{16}$O+$^{64}$Zn ~\cite{Gomes05} illustrated by the blue curve of 
Fig.~1. The $^{16}$O+$^{58}$Ni~\cite{Keeley95} as well as the 
$^{17}$F+$^{58}$Ni~\cite{Mazzocco09} total reaction cross sections (not shown 
in Fig.~1) belong also to the blue curve. The fact that the weakly-bound 
$^{17}$F and the tightly-bound $^{16}$O nuclei~\cite{Mazzocco09} show the same 
trend may indicate that nuclear structure effects still play a prominent role 
in the reaction dynamics.

\begin{figure}[htb]
\begin{minipage}[t]{80mm}
%\framebox[79mm]{\rule[-26mm]{0mm}{52mm}}
\includegraphics[width=79mm]{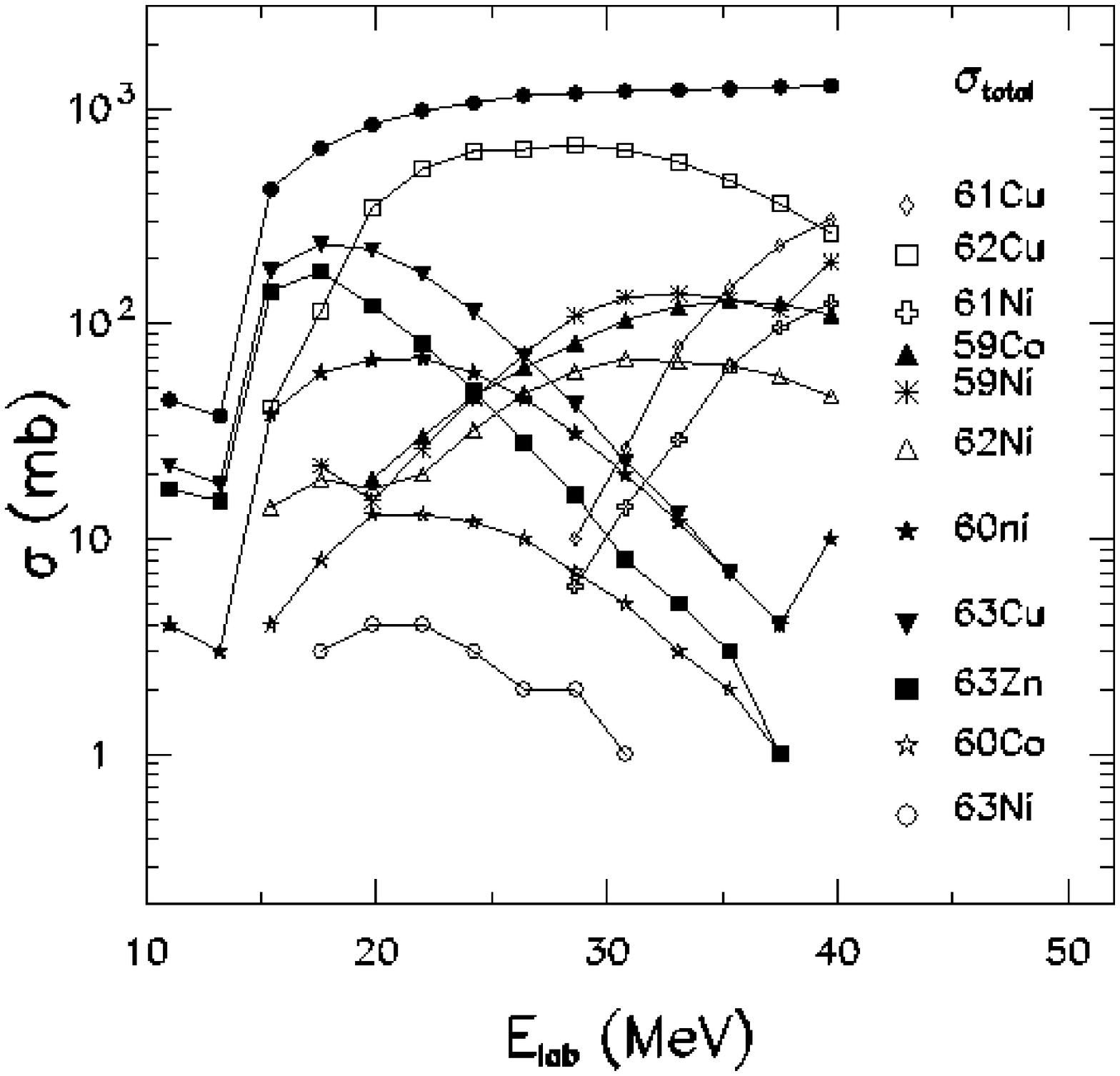}
\caption{CASCADE (CACARIZO~\cite{Mahboub}) predictions for excitation 
functions of evaporation residues produced in $^6$Li+$^{59}$Co complete fusion 
reactions. The corresponding experimental data were given in Fig.~1 of 
Ref.~\cite{Beck03}.}
\label{fig.3}
\end{minipage}
\hspace{\fill}
\begin{minipage}[t]{75mm}
%\framebox[74mm]{\rule[-26mm]{0mm}{52mm}}
\includegraphics[width=74mm]{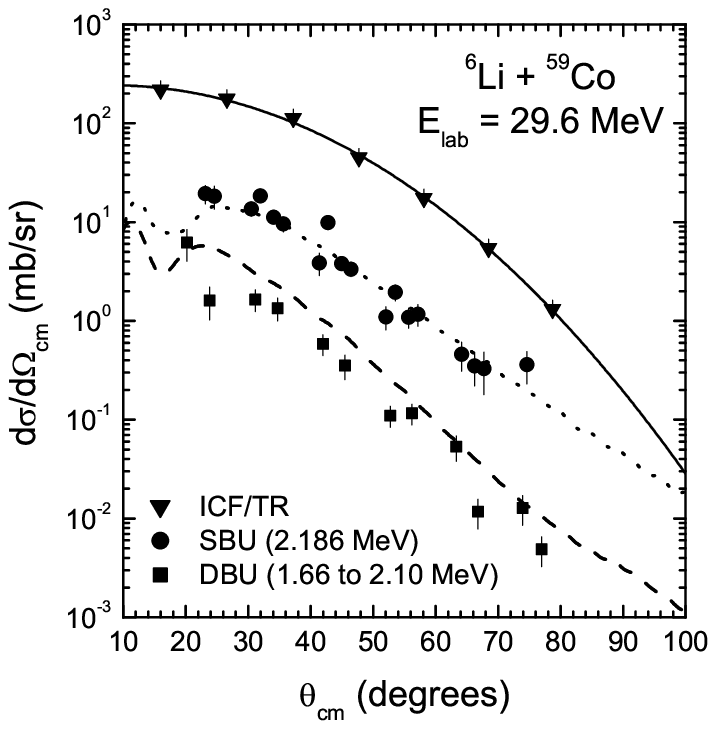}
\caption{Experimental (Full triangles circles and rectangles) and 
theoretical (solid, dashed and dotted curves) angular distributions for the 
ICF/TR, SBU and DBU processes (see text). }
\label{fig.4}
\end{minipage}
\end{figure}

Similarly, total $\alpha$ production is also found to be more intense for 
$^{6}$He+$^{64}$Zn~\cite{Pietro04} and $^{6}$He+$^{209}$Bi~\cite{Kolata98} 
when compared, in Fig.~2, to the universal function determined in 
Ref.~\cite{Souza09} for reactions induced by $^{6}$Li projectiles. It is 
interesting to note that reactions induced by $^{7}$Li 
projectiles~\cite{Pakou05} obey the same systematic trend, giving further 
support to the present comparisons. Since there still exist
contradictory results for beryllium isotopes, it will be of great interest 
to see how the recent Rex-Isolde measurements for $^{11}$Be+$^{64}$Zn
\cite{Pietro09} (for example, preliminary total reaction cross sections are 
found to be at least twice those for $^{9}$Be+$^{64}$Zn~\cite{Gomes05}) will 
follow the systematics of both  Fig.~1 and Fig.~2. For instance, Kolata and 
collaborators~\cite{Kolata04} have argued that the total reaction cross 
section measured earlier for $^{9}$Be+$^{209}$Bi \cite{Signorini00} was very 
much enhanced compared with that for $^{10}$Be+$^{208}$Pb~\cite{Kolata04} at 
sub-barrier energies, due to the weakly-bound nature of the $^{9}$Be 
projectile. However, and despite the $^{11}$Be halo structure. up to now 
no significant difference has been observed between 
$^{9,11}$Be+$^{209}$Bi \cite{Signorini00,Mazzocco06} and $^{10}$Be+$^{208}$Pb
\cite{Kolata04}. It may appear important to perform as soon as
possible new experiments of this type, but with high-quality Beryllium 
beams that have recently become available at Rex-Isolde.
 
A comprehensive study of $^{6}$Li+$^{59}$Co
\cite{Beck07b,Beck03,Souza09,Diaz03,Beck06,Beck08,Souza09b,Souza09c} 
(considered as a benchmark reaction) is still in progress. Results on total 
reaction cross sections extracted from the optical model (OM) analysis 
\cite{Souza07} of the elastic scattering (using the S\~ao Paulo Potential 
\cite{Souza07}) are shown in Fig.~1 as magenta triangles, and the 
corresponding total $\alpha$ production cross sections are indicated by full 
stars in Fig.~2. The comparison with Continuum-Discretized Coupled-Channel 
(CDCC) calculations \cite{Beck07b,Diaz03} indicates only a small enhancement 
of total fusion for the more weakly-bound $^{6}$Li below the Coulomb barrier, 
with similar cross sections for both $^{6,7}$Li+$^{59}$Co reactions at and 
above the barrier. Although rather low breakup cross sections were measured 
for $^{6,7}$Li+$^{59}$Co, even at incident energies higher than the Coulomb 
barrier \cite{Beck07b,Diaz03}, the coupling to the breakup channel is 
extremely important for the CDCC analysis \cite{Beck07b,Keeley09} of the 
elastic scattering angular distributions \cite{Souza07}.

\section{DISCUSSION}

Fig.~3 displays $^{6}$Li+$^{59}$Co excitation functions for fusion-evaporation 
residue (ER) channels as predicted by CACARIZO, the Monte Carlo version of 
CASCADE \cite{Mahboub}. This uses rather well established input parameters for 
the medium-mass region A $\approx$ 60. The unexpected disagreement with the 
experimental data for almost all of the dozen or so ER channels~\cite{Beck03} 
was interpreted as a signature of the occurence of intense incomplete-fusion 
(ICF) components. A careful investigation was later undertaken in 
Ref.~\cite{Varenna} for the three strongest ER channels for $^{6}$Li+$^{59}$Co, 
using two other well-known statistical-model codes (PACE2 and EMPIRE-II), and 
similar conclusions on the role of ICF were proposed. We would like to point 
out that the statistical-model simulation with CASCADE, presented in the 
present work, is in fairly good agreement (within 30$\%$) with both PACE2 and 
EMPIRE-II calculations~\cite{Varenna}. Although the ICF hypothesis was 
invoked, one should remain rather cautious when using evaporation codes for 
fusion induced by weakly-bound projectiles or halo projectiles such as 
$^6$He~\cite{Chatterjee08,Pietro04}. In the latter case
\cite{Chatterjee08,Pietro04}, the importance of the role of transfer channels 
rather than ICF has been proposed. However, for exotic nuclei populated
in this kind of fusion reaction, there is, unfortunately, a real lack of 
information on both OM parameters (transmission coefficients) and level 
densities (see for instance Ref.~\cite{Avrigeanu09,Oginni09}) which are among 
the key input parameters of evaporation codes.

A detailed study of the breakup process in the $^{6}$Li+$^{59}$Co reaction 
with particle techniques allowed us to discuss the interplay of fusion 
(CF and ICF) and breakup processes~\cite{Beck07b}. Coincidence data 
registered at E$_{lab}$ = 29.6 MeV, compared with three-body kinematics 
calculations~\cite{Souza09,Souza09b,Souza09c}, reveal how to disentangle the 
contributions from breakup, ICF and/or transfer-re-emission processes (TR). A 
very preliminary estimate of the total experimental ICF cross section would 
give approximately 150 mb. This value appears to be consistent with a  
calculation performed using the model of Diaz-Torres \cite{Diaz07}. 

Fig.~4 depicts the experimental angular distributions for both the 
sequential breakup (SBU) and ICF/TR processes analysed in Ref.~\cite{Souza09}, 
as well as for the direct breakup (DBU) components. For the case of ICF/TR, we 
used the differential cross sections extracted from inclusive data
\cite{Souza09}. The solid and dotted lines were extracted from Ref.
\cite{Souza09} and correspond to the ICF/TR Gaussian fit and the SBU CDCC 
calculation \cite{Beck07b}, respectively. The dashed line represents the DBU 
CDCC results \cite{Beck07b} for the $^{6}$Li excitation energy range from 
E$^*$ =  $1.48$~MeV (breakup threshold) to E$^*$ = $2.10$~MeV in the 
continuum.

\begin{figure}[htb]
\begin{minipage}[t]{80mm}
\includegraphics[width=79mm]{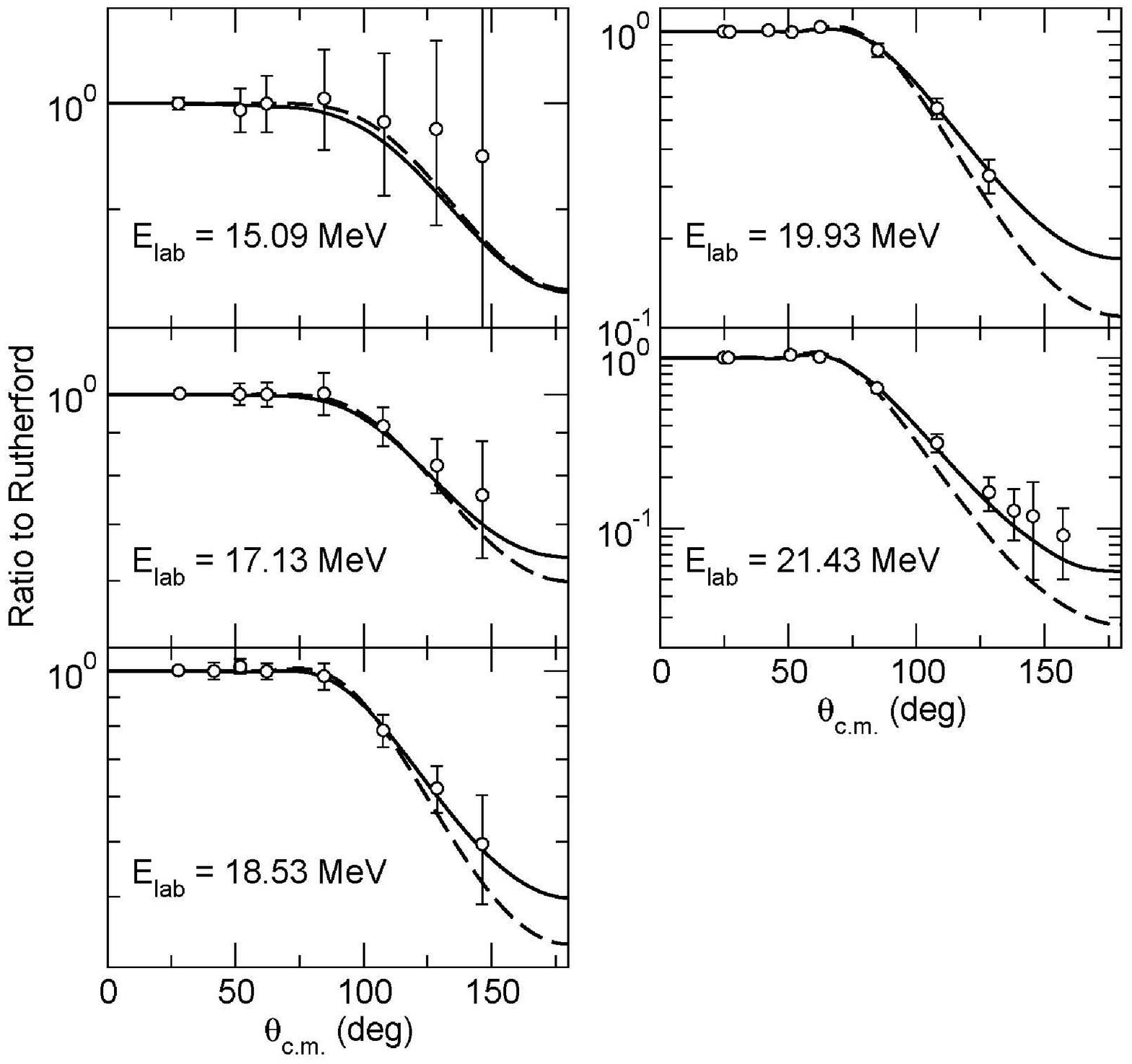}
\caption{$^7$Be+$^{58}$Ni elastic scattering measurements (experimental data 
are from \cite{Aguilera09}). Solid and dashed curves are CDCC calculations
\cite{Keeley09} with respectively full-coupling and without coupling as
explained in the text.}
\label{fig.5}
\end{minipage}
\hspace{\fill}
\begin{minipage}[t]{75mm}
\includegraphics[width=55mm]{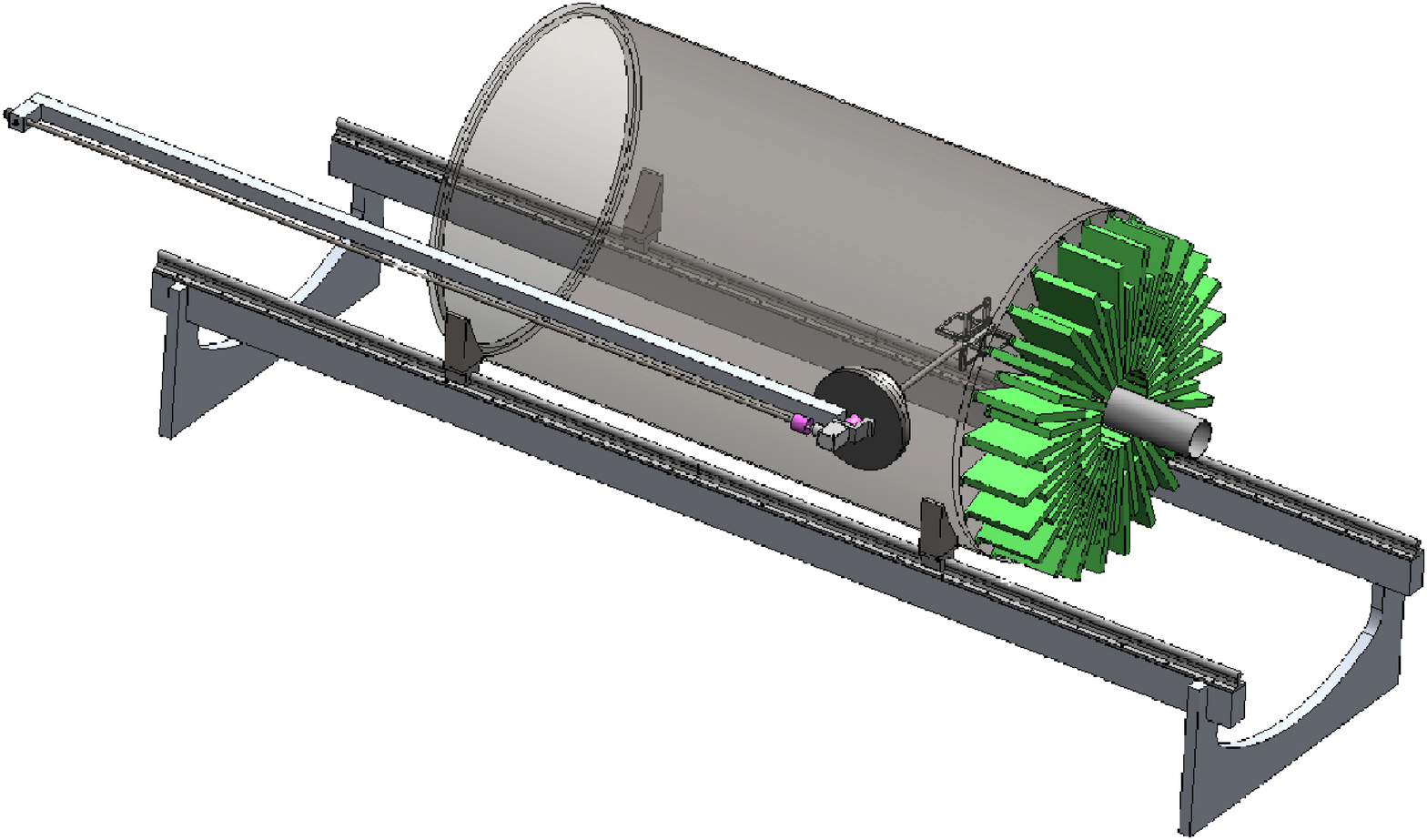}
\caption{(color online) R$\&$D design of the MSU/NSCL active-target 
time-projection chamber to be installed within the solenoid: view of the 
chamber with a removable target wheel \cite{MSU09}.}
   \label{fig.6}
\end{minipage}
\end{figure}

As far as exotic ``halo" projectiles are concerned we have initiated a 
systematic study of the $^{8}$B and $^{7}$Be induced reactions data 
\cite{Aguilera09} (partly displayed in Fig.~1) with an improved CDCC method 
\cite{Beck07b,Keeley09}. Some of the preliminary results on the angular 
distributions are displayed in Fig.~5 for the $^{7}$Be+$^{58}$Ni elastic 
scattering. As compared to $^{7}$Be+$^{58}$Ni (similar to 
$^{6,7}$Li+$^{58,60}$Ni) our CDCC analysis of the $^{8}$B+$^{58}$Ni reaction 
(not shown in the present paper since it is in full agreement with the work 
of Lubian et al. \cite{Lubian09}) while exhibiting a large breakup cross 
section (consistent with the experimental systematics \cite {Kolata09}) is 
rather surprising as regards the consequent weak-coupling effect found to be 
particularly small in the near-barrier elastic scattering measurements 
\cite{Aguilera09}. A more detailed discussion with comparisons of coupling
effects for near-barrier $^{6}$Li, $^{7}$Be and $^{8}$B elastic scattering
angular distributions is proposed in Ref.~\cite{Keeley09}. 

\section{SUMMARY, CONCLUSIONS AND OUTLOOK}

A systematic overview of the competition of various reaction mechanisms
induced by either stable, weakly-bound or exotic, halo nuclei is proposed
in Fig.~1 and Fig.~2, where a large number of recent experimental results are 
compiled and briefly presented. The correctness of the statistical-model
codes is critically discussed for reactions induced by RIB projectiles for 
which essential input parameters such as level densities and OM transmission 
coefficients are less well known. From a detailed investigation of the 
$^{6}$Li+$^{59}$Co reaction, it can be concluded that a clear separation of 
the different reaction mechanisms remains one of the main challenges in the 
study of fusion reactions induced by stable, weakly-bound and exotic, halo 
projectiles with medium-mass targets in the vicinity of the Coulomb barrier 
and below. For halo systems a full understanding of the reaction dynamics 
involving couplings to the breakup and nucleon-transfer channels will need 
high-intensity radioactive ion beams (presently available at SPIRAL/GANIL
\cite{Chatterjee08}, DRIBs/Dubna \cite{Penion06}, Rex-Isolde \cite{Pietro09}, 
MSU/Notre-Dame \cite{Aguilera09,MSU09}, RIBRAS/S\~ao Paulo \cite{Benjamin07}, 
RIPS/RIKEN \cite{Mazzocco06} ...) and precise measurements of elastic 
scattering, fusion and yields for breakup itself. A proposal \cite{MSU09} to 
study reactions such as $^{8}$B+$^{40}$Ar and $^{11}$Be+$^{40}$Ar, using an 
active-target time-projection chamber (AT-TPC), is underway at MSU/NSCL. The 
AT-TPC is a dual functionality device containing both traditional 
active-target and time-projection chamber capabilities. The detector consists 
of a large gas-filled chamber installed in an external magnetic field 
(solenoid) as shown in Fig.~6. 

\section{Acknowledgments}

{\small E.F. Aguilera and J.J. Kolata are warmly acknowledged for providing us
in several occasions with analysed data prior to their publictaion. One of us 
(CB) would like to thank N. Alamanos, S. Kailas, F. Liang, J. Lubian, M. 
Mazzocco, W. Mittig, and A. Richard for fruitful discussions during the course 
of the NN2009 Conference in Beijing.}

\end{document}